\documentclass[twocolumn,aps]{revtex4}

\usepackage{graphicx}
\usepackage{dcolumn}
\usepackage{bm}
\usepackage{epsfig}
\usepackage{citesort}
\usepackage{times}
\usepackage{float} 

\usepackage{color}
\definecolor{red}{rgb}{1,0,0}

\begin{document}

\title{Statistical mechanics of Monte Carlo sampling and the sign problem.}
\author{Gustavo D\"uring$^1$}
\author{Jorge Kurchan $^2$}

\affiliation{$^{1}$
Laboratoire de Physique Statistique, Ecole Normale Superieure, UPMC  
Univ Paris 06, UniversitŽ Paris Diderot, CNRS, 24 rue Lhomond, 75005  
Paris,  France \\
$^2$PMMH, ESPCI, 10 rue Vauquelin, CNRS UMR 7636 , Paris, France 75005}

\date{\today}

\begin{abstract}

Monte Carlo sampling of any system may be analyzed in terms of an associated glass model -- a variant of the Random Energy Model --  with, 
whenever there is a sign problem, complex fields. This model has three types of phases (liquid, frozen and  `chaotic'), as is characteristic of glass models with complex parameters.
Only  the liquid one  
yields the correct answers for the original problem, and the task is to design the simulation  to  stay inside it.
The statistical convergence of the  sampling  to the correct expectation values 
may be studied in these terms, yielding a general lower bound  for the computer  time as a function  of the free energy difference between 
the true system, and a reference one.  In this way, importance-sampling strategies may  be optimized.

\end{abstract}

\maketitle

The sign problem appears when we are asked to compute a partition function of the form:
\begin{equation}
Z= \sum_{s}  e^{- W(s)} 
\end{equation}
 and $W=W_R+iW_I$ is not real. In classical statistical mechanics  $s$ denotes  a configuration in space and $W=\beta E$ is proportional to the energy, while
 in the quantum case, $s$ are in space-time and $W$ is the quantum action (and we refer to this as `quantum Monte Carlo').
The problem is that   a stochastic  sampling cannot be performed with the complete $W$,
since then the probabilities are no longer positive and real.

 An imaginary part of $W$ arises naturally in quantum mechanics with real time, in field theories like QCD with baryons, and in some of the most  important systems in
 condensed matter. In many cases,  it is introduced  when $W$ is real and of the form $W = W_o - b \sum C_\alpha^2$ and one needs a  form linear in the $C_\alpha$:
$$Z= \sum_{s}  e^{- W_o(s)- b \sum C_\alpha^2} = \sum_s\int d{\bf \lambda} \; e^{ - W_o(s) + \sqrt{b}  \lambda_\alpha  C_\alpha - \frac{\lambda_\alpha^2}{2}}$$
The Hubbard-Stratonovich transformation above introduces an imaginary term $W_I=  \sqrt{|b|} \sum \lambda_\alpha C_\alpha$ in the repulsive case  $b<0$.
 
  {\bf Complex fields: }
Although the discussion in this paper is general, it is instructive to concentrate on a   (very large) class of systems for which the sign problem takes a particularly simple form. They are characterized by having a real action $W_R$, plus an imaginary term $W_I= i h_I M(s)$, where $M(s)$ is an integer-valued
function of the configuration, and appear in several contexts:\\
$\bullet \;\;\;\; $  $\theta $-terms in Euclidean field theories and solid state. $M(s)$ is a topological number associated with the configuration.
The case $h_I \equiv \theta= {i \pi}$, when the  partition function is real, is particularly interesting:   the term $W_I$ may change an otherwise gapped problem into
a gapless one (see e.g. \cite{Haldane,Hooft,W})\\
$\bullet \;\;\;\;$ Fermion systems such as the  Hubbard model
\begin{equation}
H =-t \sum_{<ij>,\sigma} (c^\dag_{i\sigma} c_{j \sigma} + h.c.) + U \sum_i n_{i\uparrow} n_{i\downarrow}
\end{equation}
The number operators $n_{i\uparrow},n_{i\downarrow}$ may be decoupled with a Hubbard-Stratonovich transformation (either with either continuous or
spin~\cite{Hirsch} variables, and the remaining
determinants evaluated. 
When the model is repulsive, these determinants pick up a minus sign counting  fermion world-line crossings~\cite{Loh}, and
in some cases   $M(s)$ can be shown to be  a topological number related to the Hubbard Stratonovich field $s$ \cite{Mura}.  \\
$\bullet \;\;\;\;$ An alternative strategy  to uncouple the number operators introduced by DeForcrand and Batrouni~\cite{Batrouni} uses 
a modified uncoupling with spins, and then $M(s)$ is the sum over space and time of the number of "up" spins.

In all these cases the sign problem arises because we have to calculate a partition function
 \begin{eqnarray}
 Z&=& \sum_M \; Z_M \; e^{-i h_I M}  \label{suma} \\
  Z_M &=& e^{-\beta F(M)}= \sum_s \; \delta(M(s)-M) \; e^{-W_R}  \label{suma1}
 \end{eqnarray}
 with $F(M)$ a real, extensive quantity that may be computed with standard Monte Carlo. This is a deceptively  simple expression, since the sum (\ref{suma}) contains
 detailed cancellations that require an enormous precision in the $F(M)$.
  
   {\bf Sampling:}
 In the presence of sign problem, 
 the alternative is  to use only the real part in the Monte Carlo rule, and evaluate the expectation of
a function $O$ as:
\begin{equation}
\langle O \rangle = \frac{ \langle e^{-iW_I} \; O\rangle_{\mbox{\tiny R}}}{ \langle e^{-iW_I} \rangle_{\mbox{\tiny R}}}
\end{equation}
 where $\langle \bullet \rangle_{\mbox{\tiny R}}$ denotes an average evaluated with the Boltzmann weight of the real part $W_R$,
 a quantity easily calculable with Monte Carlo sampling. 
 Somewhat more generally, we may apply reweighting,
  \begin{equation}
\langle O \rangle = \frac{ \langle e^{-A-iW_I} \; O\rangle_{\mbox{\tiny A}}}{ \langle e^{-A-iW_I} \rangle_{\mbox{\tiny A}}}
\label{expect}
\end{equation}
 where the average $\langle \bullet \rangle_{\mbox{\tiny A}}$ is associated with the  energy $-A+W_{R}$,  involving the suitably chosen  {\em importance sampling}  function $A$. 
 In practice, these expectation values are generated by  running a Monte Carlo program with energy $-A+W_R$ for a time ${\cal{T}}$. If the correlation time of the 
 Monte Carlo dynamics  is $\tau$, this is essentially  like using $ R $ independent samplings: $R={\cal{T}}/\tau$ , which we parametrize as  $R=e^{\gamma N}$.

 We may restate the problem as follows: we are given $R=e^{\gamma N}$ configurations $s_a= \{s_1,...,s_R\}$, chosen independently with probability
 \begin{equation}
P(s) = \frac{e^{A(s)-W_R(s)}} {\sum_{s'}e^{A(s')-W_R(s')}}
\label{prob}
\end{equation}
where the sum in the denominator runs over all possible values the configuration $s'$ my take.
 We  estimate averages (\ref{expect}) as:
 \begin{equation}
 \langle O \rangle_{estd.} = \frac{ \sum_{a=1}^R e^{-A(s_a)-iW_I(s_a)} O(s_a)}{ \sum_{a'=1}^R e^{-A(s_{a'})-iW_I(s_{a'})} }
\label{expect1}
\end{equation}
which is encoded in  the generating   (partition) function:
  \begin{equation}
 {\cal{Z}}(j)  = \sum_{a=1}^R e^{-A(s_a)-iW_I(s_a) + j  O(s_a)} 
\label{expect2}
\end{equation}
 Note that we use the same sampled configurations for both numerator and denominator.

{\bf Glasses:}
 The connection with glasses arises when we ask: what is the expected result of this procedure?
In order to answer this, we need to compute the expectation value of (\ref{expect1}), or, equivalently, the expectation  of the {\em logarithm} of ${\cal{Z}}$: this is a quenched average, a consequence of the random variables
appearing in both numerator and denominator of (\ref{expect1}). We calculate then:
 \begin{equation}
 \overline{\ln |{\cal{Z}}(j)|}  = \sum_{s_1,...,s_R} \;\Pi_{a=1}^{R}  \;  P(s_a) \; \;  {\ln |{\cal{Z}}_j}(s_1,...,s_R) |
\label{expect3}
\end{equation} 
 Real and imaginary parts of expectation values are obtained from:
  \begin{equation}
 \langle O \rangle_{estd} =\left. \left(\frac{\partial}{\partial j_R} - i \frac{\partial}{\partial j_I}\right) \; \overline{ \ln|{\cal{Z}}(j)|}  \right|_{j=0}
\label{gene}
\end{equation} 
We have chosen to express everything in terms of the average of the logarithm of the {\em norm} of the partition function, because 
its phase in general fluctuates from realization to realization ~\cite{Derrida1,Moukarzel}. 
 Equations  (\ref{prob}) and (\ref{expect2}) define  a disordered system, a generalization of the Random Energy Model~\cite{Derrida2}.
 The quantity ${\ln |{\cal{Z}}(j)|}$  is self-averaging in the thermodynamic limit. The result contains, as we shall see, 
 different phases, depending on the parameters of the original problem  and on the 
 number of samples $e^{\gamma N}$  : $\gamma$ is now a phase parameter. Of  these phases of the glassy model,  only one coincides with the original model, the others  give -- even as $N \rightarrow \infty$ -- wrong results for the sampling. Monte Carlo sampling in the presence of the sign problem may be understood as the art of being  in the only `good' phase.

 For clarity of presentation, we shall  now specialize to the situation described by complex fields (\ref{suma}) and (\ref{suma1}), and then mention how to proceed generally.
 We collect a number of samples $R$, classify them with their value $M$, obtained  with probability $P(M)$. The result is a set of  ${\cal{N}}(M)$ , with $\sum_M {\cal{N}}(M)=R=e^{N \gamma}$.
 Expectation values are obtained from (cfr. (\ref{expect1})):
 \begin{equation}
\langle O \rangle_{estd} = \frac{\sum_M {\cal{N}}(M) e^{i  h_i M} O(M)}{\sum_{M'} {\cal{N}}(M')e^{i  h_i M'}} \label{popo}
\end{equation}
The ${\cal{N}}$ are random variables, and we wish to calculate the expected result for  (\ref{popo}).
 We recognize a form of the  Random Energy Model (REM) with  imaginary inverse temperature $h_I$, and $M$ playing the role of energy.
   The input from the original system is through the (real) free energy densities $F(M)$, for real $M$, the quantity we estimate through our simulation.
   (The usual REM corresponds to  a quadratic $F(M)$).   The solution of this problem follows closely refs. \cite{Derrida1,Moukarzel}.
  
{ For large $N$, because of  independence, one can convince oneself that
  \begin{equation}
 \langle({\cal{N}} (M) -\langle {\cal{N}} (M) \rangle)^2 \rangle \sim\langle {\cal{N}} (M) \rangle \sim  \; R \; P(M)
\end{equation}
We have that either $R\; P(M)\ll1$, and there are no samples of that value of $M$, or   $R\;P(M) \gg 1$. In the latter case,  one has that the average 
number of samples of given $M$ is  ${\cal{N}}{(M)} \sim R P(M)$, but with an error  $\sqrt{{\cal{N}}{(M)}}$,  where:
}
\begin{equation}
 P(M) =   \frac{e^{-\beta  F(M)}}{\sum_{M'} e^{-\beta F(M')} }{\;\sim\;}e^{-\beta N [f(m)-f(m_{o})]}
 \end{equation}
{ For convenience we have defined the intensive quantities $f(m=M/N)=F(M)/N$.} The term with $f(m_o)$ comes from the normalization integral $\int dm e^{-\beta f(m)}$, and $m_o$ is the real axis saddle
    $\frac{df}{dm}\; (m_o)=0$.

 In order to calculate the result of (\ref{popo}) we start with the partition function:
    \begin{equation}
    {\cal{Z}}=\sum_M \;  \; {\cal{N}} (M) e^{-ih_I M}
   \end{equation}
Writing ${\cal{N}} \sim { R}P(M) + { R^{\frac{1}{2}}}P^{\frac{1}{2}}(M)  \eta(M)$, where $\eta(M)$ is a random number of zero mean and unit variance,
we  may split $\cal Z$ into  a deterministic {${ {\cal{Z}}}_{det} $} plus a fluctuating part {${\cal{Z}}_{err}$}. The deterministic part reads:    
\begin{eqnarray}
 { {\cal{Z}}}_{det}  &=&{ \sum_M \;  R P(M) e^{-ih_I M}}\nonumber \\
&{ \;\sim\;}& \int dm \; e^{N [{ \gamma}- \beta f(m) + \beta f(m_o) -i h_I m]}
   \label{exact}
   \end{eqnarray}
  The integral (\ref{exact})  is dominated by the saddle point 
  \begin{equation}
{ \beta}  \frac{df}{dm}(m_{sp}){+}ih_I=0
 \label{saddle}  
   \end{equation}
  Note that, unlike the real-axis saddle $m_o$, the total one $m_{sp}$ is complex. This is the correct solution, and we
  shall denote it `{ Phase I}'.
  
   The shifting of the correct saddle point into the complex plane is due to the rapid oscillations introduced by the imaginary term, 
which  have the effect of making the final result no longer dominated  by the  envelope maximum at $m_o$ -- and indeed, the final
result is exponentially smaller than the peak $e^{-N\beta f(m_o)}$.
In the case of our sampling, this means that we should also consider two effects, each one leading to a new phase:
The first is the fact that for values of $m$ distant enough from the maximum,  the sampling produces  no configurations at all.
The region  of integration is restricted to values $m$ bounded by the points $m_b$ such that ${\cal{N}}{(M)} =e^{-\beta N [f(m_b) - f(m_o)]} R \stackrel{>}{~} 1$, that is, within
the limits given by:
\begin{equation}
\beta  [f(m_b) - f(m_o)] = \gamma
\label{limit}
\end{equation}
The second effect comes from the fact that our integrand is noisy: this is particularly serious because the correct result depends upon very
strict cancellations giving an exponentially smaller answer. 
The error due to the noise of the integrand is:
 \begin{eqnarray}
{\cal{Z}}_{err}&=&    \int dm \; \sqrt{{\cal{N}}(M)} e^{- N i h_I m} \; \eta(m) \nonumber \\
&{ \;\sim\;}&
e^{{+}\frac{N\gamma}{2}}  \int dm \; e^{-N [ \beta (f(m) - \beta f(m_o) )/2 +i h_I m]} \; \eta(m) \nonumber 
   \label{noise}
   \end{eqnarray}
This is a fluctuating quantity. The expectation  of $|{\cal{Z}}_{err}|^2 $ over noise is easy to evaluate by saddle point, using $\langle \eta(m) \eta(m')\rangle \sim \delta(m-m')$, and one gets that:
\begin{equation}
|{\cal{Z}}_{err}| \sim e^{\frac{ \gamma  {N} }{2}}
\label{incoh}
\end{equation}

  All in all, there are three types of phases \cite{Derrida1,Moukarzel}, a charateristic feature of glassy systems with complex parameters: \\
  $\bullet \;\;\; $ Phase I:  The correct sampling of the original system, (up to usual   finite-size effects). Here it appears as a `liquid' phase of the glass, dominated by a  saddle point
 $m_{sp}$ that is  outside the domain of integration of (\ref{saddle}),  either because $m_{sp}$ is complex, or because it is real but beyond the original interval.\\
  $\bullet \;\;\; $  Phase II:  A `frozen'  phase, where the solution is dominated by one of the boundaries in sampling $m_{b}$ satisfying (\ref{limit}). \\
  $\bullet \;\;\; $  \ Phase III:  An `chaotic' phase: the sampling noise destroys the cancellations brought about by the oscillatory terms, and
   the peak of the `envelope' of the integrand dominates. This phase can be shown to contain a surface density of Lee-Yang zeroes in the complex plane \cite{Derrida1,Moukarzel}.
    
    In the present case, with $A=0$, phase II never dominates. The boundary between Phases I and III is given by the line where the two real parts of the free energies coincide { and when the contour deformation over the integral (\ref{exact}) passes through the saddle. Then, in order to be in the Phase I, we need}
    \begin{eqnarray}
 &{ h_I \max{(m_b)} > {\mbox{Im}}[\beta f(m_{sp})+ i h_I m_{sp}]>  h_I \min{(m_b)}}& \nonumber \\
 \label{cond1}\\
& -[ {\mbox{Re}}[\beta f(m_{sp})+{ i h_I m_{sp}}] - \beta f(m_o)] +\frac{\gamma}{2}>0 \label{cond2} &
 \label{boundary}
   \end{eqnarray}
  (note that $f(m_{sp})<f(m_o)$), which  may also be written in terms of the free energy at $h_I$ and the free energy at $h_{ I}=0$.
    {\em This is the condition for the simulation to yield the good result, which must be read as a bound  on the  simulation time through the condition that 
    $\gamma$ be large enough.} It  incorporates the well-known exponential dependence on the inverse temperature, and a free energy difference between the true model and one without the sign problem (i.e. with real field). Note that increasing the system size $N$ at fixed simulation time ${\cal{T}}$ is dangerous, as it implies
    making $\gamma$ smaller. 
     
    Consider next introducing an importance-sampling function of the form  $A=Na(m)$. Following the same reasoning as above,
    we conclude that the free energies differences now become:
    \begin{eqnarray}
    -\beta(f^I-f^{III})&=&  \beta \frac{f(m_o^+)+f({{m}}_o^-)}2  +    \frac{  a(m_o^+)-a({{m}}_o^-)}2\nonumber \\
    & &+\frac{\gamma}2 - \beta f(m_{sp}) -{ ih_I m_{sp}} \nonumber\\
-\beta (f^I-f^{II}) &=&  - \beta( f(m_{sp})-f(m_b)) -{ ih_I (m_{sp}-m_b)} \nonumber
\end{eqnarray}
     where $m_o^+$ and ${m}_o^-$ minimize $\beta f(m) \pm a(m)$, respectively, and $m_b$ are given by the generalization of (\ref{limit}):
  \begin{equation}
{\gamma} -  \beta (f(m_b)-f({{m}}_o^{{ -}})) +      a(m_b)-a({{m}}_o^{{ -}})=0 
     \end{equation}   
     Boundaries and free energy differences have changed,  and saddle may even be outside the domain of sampling between boundaries. All three phases are possible.
     
     The important point is that this phase structure, and hence the values of $\gamma$ for correct sampling,  depends on the free energy
     of the true system (which we cannot change), and on the free energies  $f(m_o^\pm)$, $f(m_b)$. The latter  are properties of the system under
     real fields, and hence easy to calculate for any $A$. One may then choose the optimal function $A$ that will minimize the $\gamma$ needed to be in phase $I$.

{\bf The general case:}  
  	It is easy to extend the  analysis of the disordered model for the general case, in which the imaginary part $W_I$ takes an abitrary form. We have to express free energies and probabilities 
    in terms of three collective variables $W_I$ and $A$, playing the role of $M$. The probability of a pair ($W_I,A$) is :
   $$P(W_I,A) \propto \sum_s \; e^{  A(s)-W_R(s)} \delta(W_I(s)-W(I)) \delta(A(s)-A) $$
   and we use it to compute the average of $\ln |{\cal {Z}}(W_I,A)|$.
    There may be again three phases, depending on whether what dominates is a saddle point of complex  ($W_I,A$), the boundary in
    the  ($W_I,A$)-plane where there is sampling, or the incoherent noise. The frontiers of these phases depend on number of samples through $\gamma$,
    and also on the choice of importance sampling function $A$.

    {\bf The Lee-Yang  zeroes:}
    At the Lee-Yang zeroes of the {\em original} problem (i..e. Phase I),  $Z=0$ and $f \rightarrow \infty$. It may seem that these are then problematic regions
    for sampling.
    Actually, this is not really so. The reason is as follows: in the large $N$ limit, Lee-Yang zeroes appear when two solutions  compete:
    \begin{equation}
    Z \sim e^{-N[f_R^1+if_I^1](h)} +    e^{-N[f_R^2+if_I^2](h)}
    \end{equation}
    (note that here we are using the Legendre-transformed free energy, in terms of $h$).
    Along the transition line in the complex plane where $f_R^1(h)=f_R^2(h)=f_R(h)$, the partition function is $Z \sim e^{-Nf_R(h)} [e^{iN f_I^1(h)} - e^{i N f_I^2(h)}]$, and  zeroes occur where    the oscillating factors cancel. In order for the oscillating factor to be of the same order
     as $N f_R(h)$, the factor $ [e^{iN f_I^1(h)} - e^{i N f_I^2(h)}]$ has to be exponentially small in $N$, and this happens only very close to each zero. We conclude that each zero has a very small region of influence in which the partition function is really small. 
\begin{figure}
\centering 
\includegraphics[width=0.75\columnwidth,trim=5mm 0mm 5mm 5mm, clip]{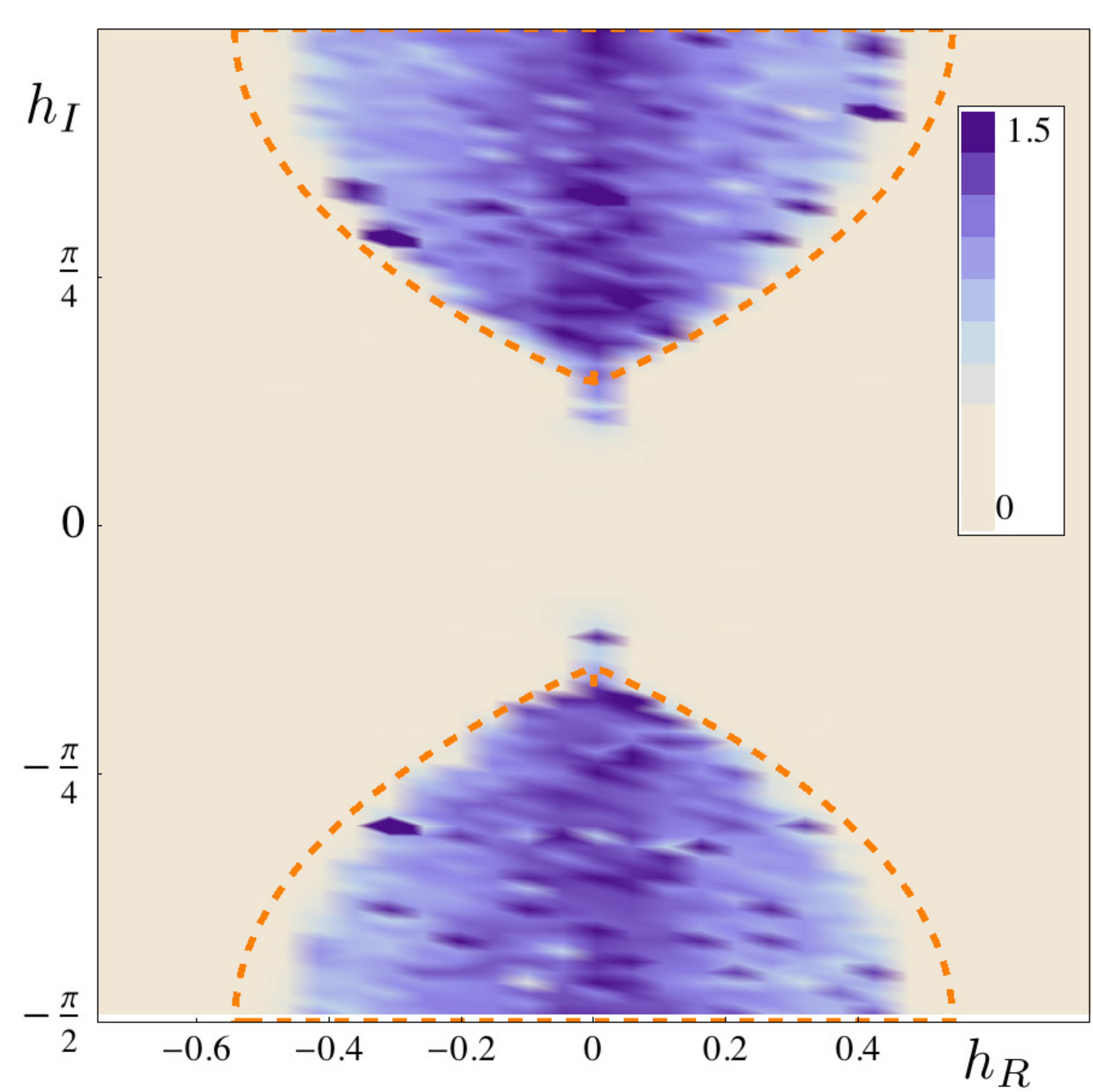}
\caption{Simulated minus exact value of the magnetization  in terms of the complex fields $h$ for $\gamma=0.5$, $J=0.5$ and $N=100$. The boundary between phases I and III is indicated with a dashed line determined through (\ref{boundary}).}
\label{error}
\end{figure} 
 
  {\bf An example: ferromagnet in a complex field:}
    Let us consider, as a simple example, a mean-field Ising ferromagnet  under a complex field $h=h_R+ih_I$. We have that
    $Z=e^{-N\beta f(m) -i h_I mN}$, with:
  \begin{eqnarray}
  -\beta f(m) &=& {-\frac{1}{2}(1+m) \ln (1+m) - \frac{1}{2}(1-m) \ln (1-m)} \nonumber \\
 & & {+\frac{\beta J}{2} m^2 - h_R m} 
  \end{eqnarray} 
  
   The correct solution of the problem is given by the saddle point equation  $m_{sp}=\tanh(\beta Jm_{sp}+h)$
   
For  complex $h$, in general $m_{sp}$ will be complex, except for $h_I=\frac{\pi}{2}$, where it will be real but outside the interval $[-1,1]$.

\begin{figure}
\centering 
\includegraphics[width=0.8\columnwidth,height=6.5cm,trim=5mm 20mm 5mm 5mm, clip]{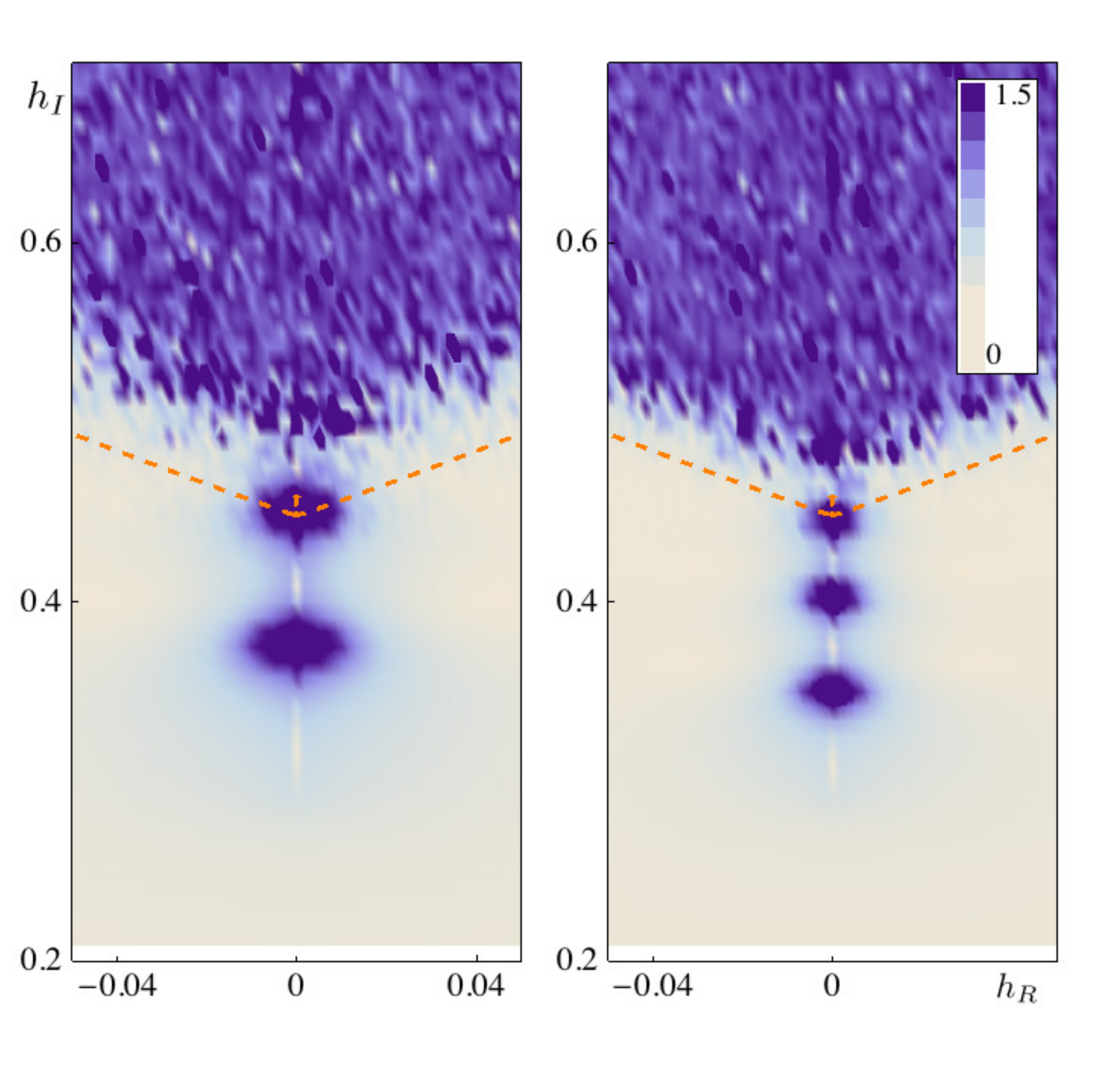}
\caption{A detail of the frontier between phases I and III { under the same conditions of Figure \ref{error}  with $N=60$ (left) and  $N=100$ (right)}. The islands visible within phase I (correct sampling) are the neighborhoods of Lee-Yang zeroes of the ferromagnet} 
\label{LY}
\end{figure}

Figure \ref{error} shows the sampled magnetization (subtracting, for clarity, the exact value) in terms of the field $h$, { with a fixed sampling time $e^{N\gamma}$}. In Figure \ref{LY} we show the frontier in 
detail, with the regions in Phase I { surrounding the lee-Yang zeroes. Clearly the size of the islands decreases for larger systems}. 

 {\bf Conclusions:}
We have shown that the error brought about by the sign problem may be discussed  in terms of the statistical mechanics of a glass model at complex
fields. A first suggestion of this analysis is that simulations should be characterized by  the size $N$ and 
 $\gamma= \frac{\ln {\cal T}}{N}$, where ${\cal T}$ is the time. A first simulation with  relatively modest values of $N$ may be used to obtain  the boundary
 $\gamma_c$ of phase $I$, and  only then  embark on larger sizes, with times determined by fixed $\gamma \stackrel{>}{\sim} \gamma_c$.

 A more sophisticated strategy is to approach the desired parameter (say, $h_I=\pi$) gradually, from smaller $h_I$, and estimate $\gamma_c(h_I)$.
 Because the sign problem is less severe for smaller values of $h_I$, one may reach the desired situation by a succession of correctly sampled 
 problems.
 
     As mentioned above, all importance sampling strategies work by increasing the free energy of phases II and III, thus making  phase I more favorable at given value of $\gamma$. Because of the nature of these phases,  this may be optimized on the basis of the knowledge of the free energy of the system at real fields alone.
            
            To conclude, let us mention that an analysis very similar to the one made here may be applied to the numerical implementation of analytic continuation.

 \end{document}